# NONLINEAR ATTITUDE STABILITY OF A SPACECRAFT ON A STATIONARY ORBIT AROUND AN ASTEROID SUBJECTED TO GRAVITY GRADIENT TORQUE[*]

Yue Wang,[†] and Shijie Xu[‡]

The classical problem of attitude stability in a central gravity field is generalized to that on a stationary orbit around a uniformly-rotating asteroid. This generalized problem is studied in the framework of geometric mechanics. Based on the natural symplectic structure, the non-canonical Hamiltonian structure of the problem is derived. The Poisson tensor, Casimir functions and equations of motion are obtained in a differential geometric method. The equilibrium of the equations of motion, i.e. the equilibrium attitude of the spacecraft, is determined from a global point of view. Nonlinear stability conditions of the equilibrium attitude are obtained with the energy-Casimir method. The nonlinear attitude stability is then investigated versus three parameters of the asteroid, including the ratio of the mean radius to the stationary orbital radius, the harmonic coefficients $C_{20}$ and $C_{22}$. It is found that when the spacecraft is located on the intermediate-moment principal axis of the asteroid, the nonlinear stability domain can be totally different from the classical Lagrange region on a circular orbit in a central gravity field.

## INTRODUCTION

Attitude stability of spacecraft subjected to the gravity gradient torque in a central gravity field has been one of the most fundamental problems in the space engineering. Attitude stability on a circular orbit in a central gravity field has been studied by Beletskii[1], DeBra and Delp[2], Hughes[3] and many other authors. Brucker and Gurfil[4] showed that the classical attitude stability domain can be modified in the restricted three-body problem by the extra primary body.

Over the last two decades, the growing interest in the scientific exploration of asteroids and the near-Earth objects (NEOs) hazard mitigation has translated into an increasing number of asteroid missions. All the major space agencies are involved on missions to NEOs and several missions are under development[5]. A thorough understanding of the dynamical behavior of spacecraft near asteroids is necessary prior to the mission design. Due to the significantly non-spherical mass distribution and the fast rotation of the asteroid, the orbital and attitude dynamics of the spacecraft are much more complex than that around the Earth. This point has been shown by many works on the orbital dynamics around asteroids, such as the works by Hirabayashi et al.[6], Hu[7], Hu and Scheeres[8], San-Juan et al.[9], Scheeres[10][11], Scheeres and Hu[12], Scheeres et al.[13][14][15], as well as by several works on the attitude dynamics around asteroids, such as the works by Kumar[16], Misra and Panchenko[17], Riverin and Misra[18], Wang and Xu[19][20][21]. Therefore, detailed investigations on the orbital and attitude dynamics of spacecraft near asteroids are of great interest and value in the future space missions to asteroids.

As shown by Kumar[16], Misra and Panchenko[17], Riverin and Misra[18], the non-central gravity field and rotational state of the asteroid disturb the attitude motion strongly. A full fourth-order model of the gravity gradient torque was derived by Wang and Xu[22] by taking into account of higher-order inertia integrals of the spacecraft. The equilibrium attitude and linear stability on a stationary orbit around an asteroid were studied by Wang and Xu[19][20] based on the linearized equations of motion. It was found that the linear stability domain was modified significantly in comparison with, even totally different from, the classical linear stability domain on a circular orbit in a central gravity field. The full nonlinear attitude dynamics on a stationary orbit around an asteroid has been analyzed via the canonical Hamiltonian formalism and the dynamical systems theory by Wang and Xu[21].

The linear attitude stability of a spacecraft on a stationary orbit around an asteroid has been studied thoroughly by Wang and Xu[20]. However, since the system is conservative and only the necessary conditions of stability can be obtained via the linearized equations of motion, the linear stability domain obtained there are only infinitesimally stable, but the stability can not be guaranteed for the finite motions. Therefore, the more practical nonlinear attitude stability, which can be guaranteed for the finite motions, needs to be investigated. In this paper, nonlinear stability of the equilibrium attitude on a stationary orbit around an asteroid is studied in the framework of geometric mechanics. As in previous works mentioned above, the harmonic coefficients $C_{20}$ and $C_{22}$ of the gravity field of the asteroid are considered in this study.

The tools of geometric mechanics have had enormous successes in many areas of mechanics[23]. Geometric mechanics has also been used in widely-ranged problems in the celestial mechanics and space engineering. Starting from the basic settings of the problem, we uncover the Lie group framework of the problem through the derivation of the Poisson tensor. Based on this Lie group framework, two powerful techniques, the determination of equilibria and the energy-Casimir method for determining nonlinear stability, are performed in this paper.

It is worth mentioning that a modified energy-Casimir method was adopted by Beck and Hall[24], and Hall[25] in the studies of attitude stability. This modified method, in which the stability problem is considered as a constrained variational problem, is more convenient for applications than the original energy-Casimir method because there is no requirement to search for a particular Casimir function. Using this modified method, we obtain the conditions of nonlinear stability. Then the nonlinear attitude stability is investigated in details in a similar manner to Reference [20] versus three important parameters of the asteroid, including the ratio of the mean radius to the stationary orbital radius, the harmonic coefficients $C_{20}$ and $C_{22}$.

**STATEMENT OF THE PROBLEM**

As described by Figure 1, we consider a rigid spacecraft $B$ moving on a stationary orbit around the asteroid $P$. The body-fixed reference frames of the asteroid and the spacecraft are defined as $S_P=\{u, v, w\}$ and $S_B=\{x, y, z\}$ with $O$ and $C$ as their origins respectively. The origin of the frame $S_P$ is at the mass center of the asteroid, and the coordinate axes are chosen to be aligned along the principal moments of inertia of the asteroid. The principal moments of inertia of the asteroid are assumed to satisfy the following inequations

$$I_{P,ww} > I_{P,vv}, I_{P,ww} > I_{P,uu}.\tag{1}$$

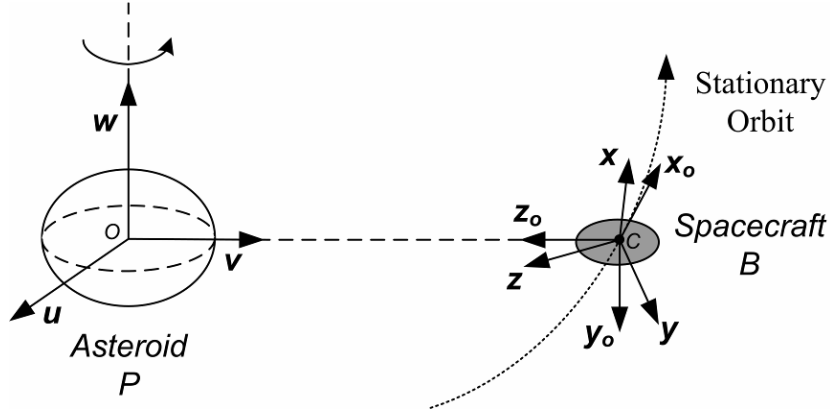

**Figure 1. The spacecraft on a stationary orbit around the asteroid.**

Then the 2nd degree and order-gravity field of the asteroid can be represented by the harmonic coefficients $C_{20}$ and $C_{22}$ with other harmonic coefficients vanished, since the origin of the frame $S_P$ is fixed at the mass center of the asteroid, and the coordinate axes are chosen to be aligned along the principal moments of inertia of the asteroid[7]. The harmonic coefficients $C_{20}$ and $C_{22}$ are defined by

$$C_{20} = -\frac{1}{2Ma_e^2}\left(2I_{P,ww} - I_{P,uu} - I_{P,vv}\right) < 0, \quad C_{22} = \frac{1}{4Ma_e^2}\left(I_{P,vv} - I_{P,uu}\right),\tag{2}$$

where $M$ and $a_e$ are the mass and the mean radius of the asteroid respectively. The ranges of the harmonic coefficients $C_{20}$ and $C_{22}$ considered in this paper are

$$-0.5 < C_{20} < 0, \quad -0.25 < C_{22} < 0.25,\tag{3}$$

which should cover most asteroids in our Solar System. The frame $S_B$ is attached to the mass center of the spacecraft and coincides with the principal axes reference frame.

We assume that the mass center of the asteroid is stationary in the inertial space, and the asteroid is in a uniform rotation around its maximum-moment principal axis, i.e. the $w$-axis. The spacecraft is on a stationary orbit, and the orbital motion is not affected by the attitude motion. According to the orbital theory by Hu[7], a stationary orbit in the inertial space corresponds to an equilibrium in the body-fixed frame of the asteroid.

There are two kinds of stationary orbits: those that lie on the intermediate-moment principal axis of the asteroid, and those that lie on the minimum-moment principal axis of the asteroid. In this paper, we assume that the spacecraft is located on the $v$-axis of the asteroid. Thus, a negative $C_{22}$ corresponds to a stationary orbit lying on the minimum-moment principal axis, and a positive $C_{22}$ corresponds to a stationary orbit lying on the intermediate-moment principal axis. According to Reference [7], the radius of the stationary orbit $R_S$ satisfies the following equation

$$R_S^5 - \frac{\mu}{\omega_T^2}\left(R_S^2 - \frac{3}{2}\tau_0 - 9\tau_2\right) = 0,\tag{4}$$

where $\mu = GM$, $G$ is the Gravitational Constant, $\tau_0 = a_e^2 C_{20}$, $\tau_2 = a_e^2 C_{22}$ and $\omega_T$ is the angular velocity of the uniform rotation of the asteroid.

As described by Figure 1, the orbital reference frame is defined by $S_o=\{x_o, y_o, z_o\}$ with its origin coinciding with $C$, the mass center of the spacecraft. $z_o$ points towards the mass center of the asteroid, $y_o$ is in the opposite direction of the orbital angular momentum, and $x_o$ completes the orthogonal triad.

**SYMPLECTIC STRUCTURE AND NON-CANONICAL HAMILTONIAN STRUCTURE**

The attitude of the spacecraft is described with respect to the orbital frame $S_o$ by $A$,

$$A = [\boldsymbol{i}, \boldsymbol{j}, \boldsymbol{k}] = [\boldsymbol{\alpha}, \boldsymbol{\beta}, \boldsymbol{\gamma}]^T \in SO(3), \tag{5}$$

where the vectors $\boldsymbol{i}, \boldsymbol{j}$ and $\boldsymbol{k}$ are components of the unit axial vectors $\boldsymbol{x}$, $\boldsymbol{y}$ and $\boldsymbol{z}$ of the frame $S_B$ in the frame $S_o$ respectively, $\boldsymbol{\alpha}$, $\boldsymbol{\beta}$ and $\boldsymbol{\gamma}$ are coordinates of the unit vectors $\boldsymbol{x}_o$, $\boldsymbol{y}_o$ and $\boldsymbol{z}_o$ in the frame $S_B$ respectively, and $SO(3)$ is the 3-dimensional special orthogonal group. The matrix $A$ is the coordinate transformation matrix from the body-fixed frame $S_B$ to the orbital frame $S_o$. Therefore, the configuration space of the problem is the Lie group

$$Q = SO(3). \tag{6}$$

The velocity phase space of the system is the tangent bundle $TQ$ with elements $(A; \dot{A})$, where $\dot{A} \in T_A SO(3)$. The image of vector $\boldsymbol{v} \in \mathbb{R}^3$ by standard isomorphism between Lie Algebras $\mathbb{R}^3$ with cross product and $so(3)$ is denoted by $\hat{\boldsymbol{v}}$, where $so(3)$ is the Lie Algebras of Lie group $SO(3)$. That is to say,

$$\hat{\boldsymbol{v}} = \begin{bmatrix} 0 & -v^3 & v^2 \\ v^3 & 0 & -v^1 \\ -v^2 & v^1 & 0 \end{bmatrix}. \tag{7}$$

The left translation of $\dot{A}$ to $so(3)$ gives

$$\hat{\boldsymbol{\Omega}}_r = T_A L_{A^{-1}} \dot{A} = A^{-1} \dot{A} \in so(3), \tag{8}$$

where $\boldsymbol{\Omega}_r$ is the relative angular velocity of the spacecraft with respect to the orbital frame $S_o$ expressed in the body-fixed frame $S_B$. We identify the tangent bundle $TSO(3)$ with $SO(3) \times \mathbb{R}^3$ with elements $(A; \boldsymbol{\Omega}_r)$ by left trivialization and the standard isomorphism $\wedge : \mathbb{R}^3 \to so(3)$. Therefore, the elements of $TQ$ can be written as $(A; \boldsymbol{\Omega}_r)$.

The angular velocity of the spacecraft with respect to the inertial space expressed in the body-fixed frame of the spacecraft $S_B$, $\boldsymbol{\Omega}$, can be calculated by

$$\boldsymbol{\Omega} = \boldsymbol{\Omega}_r + A^T \boldsymbol{\Omega}_{Orbit} = \boldsymbol{\Omega}_r + A^T \begin{bmatrix} 0 & -\omega_T & 0 \end{bmatrix}^T = \boldsymbol{\Omega}_r - \omega_T \boldsymbol{\beta}, \tag{9}$$

where $\boldsymbol{\Omega}_{Orbit}$ is the angular velocity of the orbital frame $S_o$ expressed in itself.

The rotational kinetic energy of the spacecraft is the function $T : TQ \to \mathbb{R}$ given by

$$T = \frac{1}{2} \boldsymbol{\Omega}^T \boldsymbol{I} \boldsymbol{\Omega} = \frac{1}{2} \boldsymbol{\Omega}_r^T \boldsymbol{I} \boldsymbol{\Omega}_r - \omega_T \boldsymbol{\Omega}_r^T \boldsymbol{I} \boldsymbol{\beta} + \frac{1}{2} \omega_T^2 \boldsymbol{\beta}^T \boldsymbol{I} \boldsymbol{\beta}, \tag{10}$$

where the inertia tensor $\boldsymbol{I}$ is given by

$$\boldsymbol{I} = diag\{I_{xx}, I_{yy}, I_{zz}\}, \tag{11}$$

with the principal moments of inertia of the spacecraft $I_{xx}$, $I_{yy}$ and $I_{zz}$.

The gravitational potential of the spacecraft is the function $V: Q \to \mathbb{R}$

$$V = V(\boldsymbol{\alpha}, \boldsymbol{\beta}, \boldsymbol{\gamma}). \tag{12}$$

According to the results by Wang and Xu[19], due to the significantly non-spherical shape and the rapid rotation of the asteroid, the effects of the harmonic coefficients $C_{20}$ and $C_{22}$ are as significant as that of the central component of the gravity field of the asteroid, while effects of the third and fourth-order inertia integrals of the spacecraft could be neglected. Therefore, we only consider the moments of inertia $I_{xx}$, $I_{yy}$ and $I_{zz}$ in the gravitational potential, with the third and fourth-order inertia integrals of the spacecraft neglected. Based on the results by Wang and Xu[22], through some rearrangements, the explicit formulation of the attitude-dependent part of the gravitational potential $V(\boldsymbol{\alpha}, \boldsymbol{\beta}, \boldsymbol{\gamma})$ is given by

$$V(\boldsymbol{\alpha}, \boldsymbol{\beta}, \boldsymbol{\gamma}) = \frac{3\mu}{2R_S^3}\boldsymbol{\gamma}^T\boldsymbol{I}\boldsymbol{\gamma} + \frac{3\mu\tau_0}{2R_S^5}\left(\boldsymbol{\beta}^T\boldsymbol{I}\boldsymbol{\beta} - \frac{5}{2}\boldsymbol{\gamma}^T\boldsymbol{I}\boldsymbol{\gamma}\right) - \frac{3\mu\tau_2}{2R_S^5}\left(17\boldsymbol{\gamma}^T\boldsymbol{I}\boldsymbol{\gamma} - 2\boldsymbol{\alpha}^T\boldsymbol{I}\boldsymbol{\alpha}\right). \tag{13}$$

Then, the Lagrangian of the system $L: TQ \to \mathbb{R}$ is given as follows:

$$L = T - V \circ \tau, \tag{14}$$

where $\tau: TQ \to Q$ is the canonical projection.

The (momentum) phase space is the cotangent bundle $T^*Q$, which can be written as $(A; \boldsymbol{\alpha}_A)$ with $\boldsymbol{\alpha}_A \in T_A^*SO(3)$. By left trivialization, the elements of $T^*Q$ can be written as $(A; \boldsymbol{\alpha})$ with $\boldsymbol{\alpha} = T_e^*L_A\boldsymbol{\alpha}_A \in so(3)^*$, where $so(3)^*$ is the dual space to Lie Algebra $so(3)$. We identify $so(3)^*$ with $\mathbb{R}^3$ using the standard isomorphism $\wedge$, and the pairing between $so(3)^*$ and $so(3)$ is defined as the dot product on $\mathbb{R}^3$

$$\langle \boldsymbol{a}, \boldsymbol{b} \rangle = \boldsymbol{a} \cdot \boldsymbol{b} = \frac{1}{2}tr(\hat{\boldsymbol{a}}^T\hat{\boldsymbol{b}}). \tag{15}$$

This pairing is extended to $T^*SO(3)$ and $TSO(3)$ by left translation as follows:

$$\langle \boldsymbol{a}_A, \boldsymbol{b}_A \rangle = \frac{1}{2}tr(\boldsymbol{a}_A^T\boldsymbol{b}_A) = \frac{1}{2}tr(\hat{\boldsymbol{a}}^T\hat{\boldsymbol{b}}) = \boldsymbol{a} \cdot \boldsymbol{b}, \tag{16}$$

where $\hat{\boldsymbol{a}} = T_e^*L_A\boldsymbol{a}_A \in so(3)^*$ and $\hat{\boldsymbol{b}} = T_A L_{A^{-1}}\boldsymbol{b}_A \in so(3)$.

By the means of Legendre transformation, we can obtain the conjugate momenta as follows:

$$\boldsymbol{\Pi} = \frac{\partial L}{\partial \boldsymbol{\Omega}_r} = \boldsymbol{I}\boldsymbol{\Omega}_r - \omega_T \boldsymbol{I}\boldsymbol{\beta}, \tag{17}$$

where $\partial f / \partial \boldsymbol{v}$ represents the gradient of the function $f$ with respect to the vector $\boldsymbol{v}$. $\boldsymbol{\Pi}$ is the angular momentum of the spacecraft with respect to the inertial frame expressed in the body-fixed frame $S_B$.

Since $\boldsymbol{\Omega}_r \in so(3)$ and $\boldsymbol{\Pi} \in so(3)^*$, the pairing between them can be written as

$$\boldsymbol{\Pi} \cdot \boldsymbol{\Omega}_r = \frac{1}{2} tr(\hat{\boldsymbol{\Pi}}^T \hat{\boldsymbol{\Omega}}_r) = \frac{1}{2} tr(\hat{\boldsymbol{\Pi}}^T \boldsymbol{A}^{-1} \dot{\boldsymbol{A}}) = \frac{1}{2} tr\left(\left(\boldsymbol{A}\hat{\boldsymbol{\Pi}}\right)^T \dot{\boldsymbol{A}}\right). \tag{18}$$

Since $\dot{\boldsymbol{A}} \in T_A SO(3)$, using Eqs. (16) and (18) we can obtain

$$\boldsymbol{A}\hat{\boldsymbol{\Pi}} = T_A^* L_{A^{-1}} \hat{\boldsymbol{\Pi}} \in T_A^* SO(3). \tag{19}$$

Therefore, elements of the phase space $T^*Q$ can be written as

$$\boldsymbol{\Xi} = (\boldsymbol{A}; \boldsymbol{A}\hat{\boldsymbol{\Pi}}). \tag{20}$$

The phase space $T^*Q$ carries a natural symplectic structure $\omega$, defined as

$$\omega = \omega^{SO(3)}. \tag{21}$$

The canonical bracket associated to the symplectic structure $\omega$ can be written in the coordinates $\boldsymbol{\Xi}$ as follows:

$$\{f, g\}_{T^*Q}(\boldsymbol{\Xi}) = \langle D_A f, D_{A\hat{\Pi}} g \rangle - \langle D_A g, D_{A\hat{\Pi}} f \rangle \tag{22}$$

for any $f, g \in C^\infty(T^*Q)$, where $\langle \cdot, \cdot \rangle$ is the pairing between $T^*SO(3)$ and $TSO(3)$, and $D_B f$ is a matrix whose elements are the partial derivates of $f$ with respect to the elements of matrix $\boldsymbol{B}$ respectively.

By the means of Legendre transformation, the Hamiltonian of the system $H: T^*Q \to \mathbb{R}$ is obtained as follows:

$$H = \frac{1}{2} \boldsymbol{\Pi}^T \boldsymbol{I}^{-1} \boldsymbol{\Pi} - \boldsymbol{\Pi}^T \left(-\omega_T \boldsymbol{\beta}\right) + V \circ \tau_{T*Q}, \tag{23}$$

where $\tau_{T*Q} : T^*Q \to Q$ is the canonical projection. We can see that the first term in Eq. (23) is the Hamiltonian of the free-spin motion, and the attitude dynamics of the spacecraft is perturbed both by the second and third terms of the Hamiltonian in Eq. (23). The third term $V$ represents the perturbation due to the gravity gradient torque; the second term $-\boldsymbol{\Pi}^T\left(-\omega_T \boldsymbol{\beta}\right)$ represents the perturbation due to the precession of the orbital frame, which is consistent with the results by Gurfil et al.[26], Wang and Xu[21].

Although the coordinates $\boldsymbol{\Xi}$ in Eq. (20), the symplectic structure $\omega$ in Eq. (21) and the canonical bracket Eq. (22) are natural and intrinsic, they are not convenient for applications, since the variables for the attitude motion are given in the matrix form and the calculations of the pairing between $T^*SO(3)$ and $TSO(3)$ in Eq. (22) are tedious. The non-canonical Hamiltonian structure with variables in the vector form is more convenient for applications.

We can choose a set of coordinates of the phase space $T^*Q$ instead of $\boldsymbol{\Xi}$ as

$$\boldsymbol{z} = \left[\boldsymbol{\Pi}^T, \boldsymbol{\alpha}^T, \boldsymbol{\beta}^T, \boldsymbol{\gamma}^T\right]^T \in \mathbb{R}^{12}. \tag{24}$$

There exists a Poisson diffeomorphism $\Psi : \left(T^*Q, \{\cdot, \cdot\}_{T^*Q}(\boldsymbol{\Xi})\right) \to \left(\mathbb{R}^{12}, \{\cdot, \cdot\}_{\mathbb{R}^{12}}(\boldsymbol{z})\right)$, defined as follows:

$$\Psi(A; A\hat{\boldsymbol{\Pi}}) = \left[\boldsymbol{\Pi}^T, \boldsymbol{\alpha}^T, \boldsymbol{\beta}^T, \boldsymbol{\gamma}^T\right]^T, \qquad (25)$$

where $\{\cdot,\cdot\}_{\mathbb{R}^{12}}(z)$ is the Poisson bracket in coordinates $z$. These two brackets satisfy

$$\{f, g\}_{\mathbb{R}^{12}}(z) \circ \Psi = \{f \circ \Psi, g \circ \Psi\}_{T^*Q}(\boldsymbol{\Xi}) \qquad (26)$$

for any $f, g \in C^{\infty}(\mathbb{R}^{12})$. We write Poisson bracket $\{\cdot,\cdot\}_{\mathbb{R}^{12}}(z)$ in the following form

$$\{f, g\}_{\mathbb{R}^{12}}(z) = (\nabla_z f)^T B(z)(\nabla_z g), \qquad (27)$$

with the Poisson tensor $B(z)$ (see Reference [27] for the derivation of $B(z)$) given by

$$B(z) = \begin{bmatrix} \hat{\boldsymbol{\Pi}} & \hat{\boldsymbol{\alpha}} & \hat{\boldsymbol{\beta}} & \hat{\boldsymbol{\gamma}} \\ \hat{\boldsymbol{\alpha}} & 0 & 0 & 0 \\ \hat{\boldsymbol{\beta}} & 0 & 0 & 0 \\ \hat{\boldsymbol{\gamma}} & 0 & 0 & 0 \end{bmatrix}. \qquad (28)$$

The $12\times 12$ antisymmetric and degenerated Poisson tensor $B(z)$ has six geometric integrals as independent Casimir functions

$$C_1(z) = \frac{1}{2}\boldsymbol{\alpha}^T\boldsymbol{\alpha}, \; C_2(z) = \frac{1}{2}\boldsymbol{\beta}^T\boldsymbol{\beta}, \; C_3(z) = \frac{1}{2}\boldsymbol{\gamma}^T\boldsymbol{\gamma}, \; C_4(z) = \boldsymbol{\alpha}^T\boldsymbol{\beta}, \; C_5(z) = \boldsymbol{\alpha}^T\boldsymbol{\gamma}, \; C_6(z) = \boldsymbol{\beta}^T\boldsymbol{\gamma}.$$

The six-dimensional invariant manifold or symplectic leaf of the system can be defined in $\mathbb{R}^{12}$ by Casimir functions

$$\Sigma = \left\{ z \in \mathbb{R}^{12} \mid C_1(z) = C_2(z) = C_3(z) = \frac{1}{2}, C_4(z) = C_5(z) = C_6(z) = 0 \right\}. \qquad (29)$$

The symplectic structure on this symplectic leaf is defined by restriction of the Poisson bracket $\{\cdot,\cdot\}_{\mathbb{R}^{12}}(z)$ to $\Sigma$. The six-dimensional nullspace of $B(z)$ can be got from Casimir functions

$$\mathrm{N}[B(z)] = \mathrm{span}\left\{ \begin{pmatrix} 0 \\ \boldsymbol{\alpha} \\ 0 \\ 0 \end{pmatrix}, \begin{pmatrix} 0 \\ 0 \\ \boldsymbol{\beta} \\ 0 \end{pmatrix}, \begin{pmatrix} 0 \\ 0 \\ 0 \\ \boldsymbol{\gamma} \end{pmatrix}, \begin{pmatrix} 0 \\ \boldsymbol{\beta} \\ \boldsymbol{\alpha} \\ 0 \end{pmatrix}, \begin{pmatrix} 0 \\ \boldsymbol{\gamma} \\ 0 \\ \boldsymbol{\alpha} \end{pmatrix}, \begin{pmatrix} 0 \\ 0 \\ \boldsymbol{\gamma} \\ \boldsymbol{\beta} \end{pmatrix} \right\}. \qquad (30)$$

With Hamiltonian in coordinates $z$ given by Eq. (23), equations of motion can be written as

$$\dot{z} = B(z)\nabla_z H(z). \qquad (31)$$

Explicit equations of the motion can be obtained from Eqs. (23) and (31) as follows:

$$\begin{bmatrix} \dot{\boldsymbol{\Pi}} \\ \dot{\boldsymbol{\alpha}} \\ \dot{\boldsymbol{\beta}} \\ \dot{\boldsymbol{\gamma}} \end{bmatrix} = B(z) \begin{bmatrix} \boldsymbol{I}^{-1}\boldsymbol{\Pi} + \omega_T \boldsymbol{\beta} \\ \partial V/\partial\boldsymbol{\alpha} \\ \partial V/\partial\boldsymbol{\beta} + \omega_T \boldsymbol{\Pi} \\ \partial V/\partial\boldsymbol{\gamma} \end{bmatrix} = \begin{bmatrix} \hat{\boldsymbol{\Pi}}\, \boldsymbol{I}^{-1}\boldsymbol{\Pi} + \hat{\boldsymbol{\alpha}}(\partial V/\partial\boldsymbol{\alpha}) + \hat{\boldsymbol{\beta}}(\partial V/\partial\boldsymbol{\beta}) + \hat{\boldsymbol{\gamma}}(\partial V/\partial\boldsymbol{\gamma}) \\ \hat{\boldsymbol{\alpha}}\left(\boldsymbol{I}^{-1}\boldsymbol{\Pi} + \omega_T \boldsymbol{\beta}\right) \\ \hat{\boldsymbol{\beta}}\left(\boldsymbol{I}^{-1}\boldsymbol{\Pi}\right) \\ \hat{\boldsymbol{\gamma}}\left(\boldsymbol{I}^{-1}\boldsymbol{\Pi} + \omega_T \boldsymbol{\beta}\right) \end{bmatrix}. \qquad (32)$$

The term $\hat{\boldsymbol{\alpha}}(\partial V/\partial \boldsymbol{\alpha}) + \hat{\boldsymbol{\beta}}(\partial V/\partial \boldsymbol{\beta}) + \hat{\boldsymbol{\gamma}}(\partial V/\partial \boldsymbol{\gamma})$ in Eq. (32) is actually the gravity gradient torque of the spacecraft $\boldsymbol{T}_B$ expressed in the body-fixed frame $S_B$, the explicit formulation of which is given as follows:

$$\boldsymbol{T}_B = \frac{3\mu}{R_S^3}\hat{\boldsymbol{\gamma}}\boldsymbol{I}\boldsymbol{\gamma} + \frac{3\mu\tau_0}{R_S^5}\left(\hat{\boldsymbol{\beta}}\boldsymbol{I}\boldsymbol{\beta} - \frac{5}{2}\hat{\boldsymbol{\gamma}}\boldsymbol{I}\boldsymbol{\gamma}\right) - \frac{3\mu\tau_2}{R_S^5}(17\hat{\boldsymbol{\gamma}}\boldsymbol{I}\boldsymbol{\gamma} - 2\hat{\boldsymbol{\alpha}}\boldsymbol{I}\boldsymbol{\alpha}). \tag{33}$$

## EQUILIBRIUM ATTITUDE AND CONDITIONS OF NONLINEAR STABILITY

### Equilibrium Attitude

The equilibrium attitude of the spacecraft corresponds to a stationary point of the Hamiltonian constrained by the Casimir functions. The stationary point can be determined by the first variation conditions of variational Lagrangian $\nabla F(\boldsymbol{z}_e) = \boldsymbol{0}$. The variational Lagrangian $F(\boldsymbol{z})$ is given by

$$F(\boldsymbol{z}) = H(\boldsymbol{z}) - \sum_{i=1}^{6} \mu_i C_i(\boldsymbol{z}), \tag{34}$$

where the subscript $e$ is used to denote the value at the equilibria.

By using the formulations of the Hamiltonian and the Casimir functions, the equilibrium conditions are obtained as follows:

$$\boldsymbol{I}^{-1}\boldsymbol{\Pi}_e + \omega_T \boldsymbol{\beta}_e = \boldsymbol{0}, \tag{35a}$$

$$\frac{6\mu\tau_2}{R_S^5}\boldsymbol{I}\boldsymbol{\alpha}_e - \mu_1 \boldsymbol{\alpha}_e - \mu_4 \boldsymbol{\beta}_e - \mu_5 \boldsymbol{\gamma}_e = \boldsymbol{0}, \tag{35b}$$

$$\omega_T \boldsymbol{\Pi}_e + \frac{3\mu\tau_0}{R_S^5}\boldsymbol{I}\boldsymbol{\beta}_e - \mu_2 \boldsymbol{\beta}_e - \mu_4 \boldsymbol{\alpha}_e - \mu_6 \boldsymbol{\gamma}_e = \boldsymbol{0}, \tag{35c}$$

$$\left(\frac{3\mu}{R_S^3} - \frac{15\mu\tau_0}{2R_S^5} - \frac{51\mu\tau_2}{R_S^5}\right)\boldsymbol{I}\boldsymbol{\gamma}_e - \mu_3 \boldsymbol{\gamma}_e - \mu_5 \boldsymbol{\alpha}_e - \mu_6 \boldsymbol{\beta}_e = \boldsymbol{0}. \tag{35d}$$

Eq. (35a) implies that

$$\boldsymbol{\Pi}_e = -\omega_T \boldsymbol{I}\boldsymbol{\beta}_e, \tag{36}$$

and the spacecraft is stationary with respect to the orbital frame, i.e. this solution is an equilibrium attitude of the spacecraft.

Taking the dot product of $\boldsymbol{\beta}_e$ with Eq. (35b) yields $\mu_4 = (6\mu\tau_2/R_S^5)\boldsymbol{\beta}_e^T \boldsymbol{I}\boldsymbol{\alpha}_e$, while the dot product of $\boldsymbol{\alpha}_e$ with Eq. (35c) yields $\mu_4 = (3\mu\tau_0/R_S^5 - \omega_T^2)\boldsymbol{\alpha}_e^T \boldsymbol{I}\boldsymbol{\beta}_e$. Here we consider a general case

$$\frac{3\mu\tau_0}{R_S^5} - \omega_T^2 \neq \frac{6\mu\tau_2}{R_S^5}, \tag{37}$$

by Eq. (4) which is equivalent to

$$\left(\frac{R_s}{a_e}\right)^2 \neq \frac{9}{2}C_{20} + 3C_{22}. \tag{38}$$

In this case we have $\mu_4 = 0$ due to the symmetry of the inertia tensor. In the same method, we will have $\mu_5 = 0$ and $\mu_6 = 0$ in the general case when

$$\left(\frac{R_s}{a_e}\right)^2 \neq \frac{5}{2}C_{20} + 19C_{22}, \quad \left(\frac{R_s}{a_e}\right)^2 \neq 3C_{20} + 15C_{22} \tag{39}$$

respectively. Then the equilibrium conditions (35b)-(35d) imply that $\boldsymbol{\alpha}_e$, $\boldsymbol{\beta}_e$ and $\boldsymbol{\gamma}_e$ must be principal axes of the inertial tensor, i.e. the orbital frame is parallel to the body-fixed frame. This contains 24 equilibrium attitudes, only one of which is found by Wang and Xu[19] in the linear method, since geometric mechanics adopted here allows the determination of the equilibrium attitude from a global point of view.

**Conditions of Nonlinear Stability**

Without of loss of generality, we choose one of the equilibrium attitudes as follows for stability conditions

$$\boldsymbol{\Pi}_e = [0, -\omega_T I_{yy}, 0]^T, \boldsymbol{\alpha}_e = [1, 0, 0]^T, \boldsymbol{\beta}_e = [0, 1, 0]^T, \boldsymbol{\gamma}_e = [0, 0, 1]^T, \tag{40a}$$

$$\mu_1 = \frac{6\mu\tau_2}{R_S^5}I_{xx}, \mu_2 = \left(\frac{3\mu\tau_0}{R_S^5} - \omega_T^2\right)I_{yy}, \mu_3 = \left(\frac{3\mu}{R_S^3} - \frac{15\mu\tau_0}{2R_S^5} - \frac{51\mu\tau_2}{R_S^5}\right)I_{zz}. \tag{40b}$$

Following the energy-Casimir method adopted by Beck and Hall[24], and Hall[25], we can obtain the conditions of nonlinear stability. The Hessian of the variational Lagrangian is calculated:

$$\nabla^2 F(z) = \begin{bmatrix} \boldsymbol{I}^{-1} & 0 & \omega_T \mathbf{I}_{3\times 3} & 0 \\ 0 & \frac{6\mu\tau_2}{R_S^5}\boldsymbol{I} - \mu_1 \mathbf{I}_{3\times 3} & -\mu_4 \mathbf{I}_{3\times 3} & -\mu_5 \mathbf{I}_{3\times 3} \\ \omega_T \mathbf{I}_{3\times 3} & -\mu_4 \mathbf{I}_{3\times 3} & \frac{3\mu\tau_0}{R_S^5}\boldsymbol{I} - \mu_2 \mathbf{I}_{3\times 3} & -\mu_6 \mathbf{I}_{3\times 3} \\ 0 & -\mu_5 \mathbf{I}_{3\times 3} & -\mu_6 \mathbf{I}_{3\times 3} & \left(\frac{3\mu}{R_S^3} - \frac{15\mu\tau_0}{2R_S^5} - \frac{51\mu\tau_2}{R_S^5}\right)\boldsymbol{I} - \mu_3 \mathbf{I}_{3\times 3} \end{bmatrix}, \tag{41}$$

where $\mathbf{I}_{3\times 3}$ is the is the $3\times 3$ identity matrix. At the equilibrium attitude, we have

$$\nabla^2 F(z_e) = \begin{bmatrix} \boldsymbol{I}^{-1} & 0 & \omega_T \mathbf{I}_{3\times 3} & 0 \\ 0 & \frac{6\mu\tau_2}{R_S^5}(\boldsymbol{I} - I_{xx}\mathbf{I}_{3\times 3}) & 0 & 0 \\ \omega_T \mathbf{I}_{3\times 3} & 0 & \frac{3\mu\tau_0}{R_S^5}(\boldsymbol{I} - I_{yy}\mathbf{I}_{3\times 3}) + \omega_T^2 I_{yy}\mathbf{I}_{3\times 3} & 0 \\ 0 & 0 & 0 & \left(\frac{3\mu}{R_S^3} - \frac{15\mu\tau_0}{2R_S^5} - \frac{51\mu\tau_2}{R_S^5}\right) \times (\boldsymbol{I} - I_{zz}\mathbf{I}_{3\times 3}) \end{bmatrix}. \tag{42}$$

The Hamiltonian system is non-canonical, and the phase flow of the system is constrained on the six-dimensional invariant manifold or symplectic leaf $\Sigma$. Therefore, rather than considering general perturbations in the phase space, we need to restrict consideration to perturbations on $T\Sigma|z_e$, the tangent space to the invariant manifold $\Sigma$ at the equilibrium, i.e. the range space the Poisson tensor $\boldsymbol{B}(z)$ at the equilibrium, denoted by $\mathrm{R}(\boldsymbol{B}(z_e))$.

According to the results by Beck and Hall[24], the projected Hessian matrix is given by $\boldsymbol{P}(z_e)\nabla^2 F(z_e)\boldsymbol{P}(z_e)$, where the projection operator is given by

$$\boldsymbol{P}(z_e) = \boldsymbol{I}_{12\times 12} - \boldsymbol{K}(z_e)\left(\boldsymbol{K}(z_e)^T \boldsymbol{K}(z_e)\right)^{-1} \boldsymbol{K}(z_e)^T, \tag{43}$$

$$\boldsymbol{K}(z_e) = \mathrm{N}[\boldsymbol{B}(z_e)] = \begin{bmatrix} 0 & 0 & 0 & 0 & 0 & 0 \\ \alpha_e & 0 & 0 & \beta_e & \gamma_e & 0 \\ 0 & \beta_e & 0 & \alpha_e & 0 & \gamma_e \\ 0 & 0 & \gamma_e & 0 & \alpha_e & \beta_e \end{bmatrix}. \tag{44}$$

The 12×12 projected Hessian matrix will have six zero eigenvalues associated with the six-dimensional nullspace $\mathrm{N}[\boldsymbol{B}(z_e)]$, i.e. the complement space of the tangent space to the invariant manifold at the equilibrium. The remaining six eigenvalues are associated with the six-dimensional tangent space to the invariant manifold $T\Sigma|z_e$. If they are all positive, $z_e$ is a constrained minimum on the invariant manifold $\Sigma$ and the equilibrium attitude is nonlinear stability. Through some calculations, we get the projected Hessian matrix as follows:

$$\boldsymbol{P}(z_e)\nabla^2 F(z_e)\boldsymbol{P}(z_e) =$$

$$\begin{bmatrix}
\frac{1}{I_{xx}} & 0 & 0 & 0 & -\frac{1}{2}\omega_T & 0 & \frac{1}{2}\omega_T & 0 & 0 & 0 & 0 & 0 \\
0 & \frac{1}{I_{yy}} & 0 & 0 & 0 & 0 & 0 & 0 & 0 & 0 & 0 & 0 \\
0 & 0 & \frac{1}{I_{zz}} & 0 & 0 & 0 & 0 & 0 & \frac{1}{2}\omega_T & 0 & -\frac{1}{2}\omega_T & 0 \\
0 & 0 & 0 & 0 & 0 & 0 & 0 & 0 & 0 & 0 & 0 & 0 \\
-\frac{1}{2}\omega_T & 0 & 0 & 0 & M_1 & 0 & -M_1 & 0 & 0 & 0 & 0 & 0 \\
0 & 0 & 0 & 0 & 0 & M_2 & 0 & 0 & 0 & -M_2 & 0 & 0 \\
\frac{1}{2}\omega_T & 0 & 0 & 0 & -M_1 & 0 & M_1 & 0 & 0 & 0 & 0 & 0 \\
0 & 0 & 0 & 0 & 0 & 0 & 0 & 0 & 0 & 0 & 0 & 0 \\
0 & 0 & \frac{1}{2}\omega_T & 0 & 0 & 0 & 0 & 0 & M_3 & 0 & -M_3 & 0 \\
0 & 0 & 0 & 0 & 0 & -M_2 & 0 & 0 & 0 & M_2 & 0 & 0 \\
0 & 0 & -\frac{1}{2}\omega_T & 0 & 0 & 0 & 0 & 0 & -M_3 & 0 & M_3 & 0 \\
0 & 0 & 0 & 0 & 0 & 0 & 0 & 0 & 0 & 0 & 0 & 0
\end{bmatrix}, \tag{45}$$

where

$$M_1 = \frac{1}{4}\frac{6\mu\tau_2}{R_S^5}(I_{yy}-I_{xx}) - \frac{1}{4}\frac{3\mu\tau_0}{R_S^5}(I_{yy}-I_{xx}) + \frac{1}{4}\omega_T^2 I_{yy}, \tag{46}$$

$$M_2 = \frac{1}{4}\frac{6\mu\tau_2}{R_S^5}(I_{zz}-I_{xx}) - \frac{1}{4}\left(\frac{3\mu}{R_S^3}-\frac{15\mu\tau_0}{2R_S^5}-\frac{51\mu\tau_2}{R_S^5}\right)(I_{zz}-I_{xx}), \tag{47}$$

$$M_3 = \frac{1}{4}\frac{3\mu\tau_0}{R_S^5}(I_{zz}-I_{yy}) - \frac{1}{4}\left(\frac{3\mu}{R_S^3}-\frac{15\mu\tau_0}{2R_S^5}-\frac{51\mu\tau_2}{R_S^5}\right)(I_{zz}-I_{yy}) + \frac{1}{4}\omega_T^2 I_{yy}. \tag{48}$$

The eigenvalues of the projected Hessian matrix are calculated as follows:

$$\{0, 0, 0, 0, 0, 0, 1/I_{yy}, 2M_2, \sigma_1, \sigma_2, \sigma_3, \sigma_4\}, \tag{49}$$

where

$$\sigma_1 = \frac{1}{2I_{xx}}\left\{1+2M_1 I_{xx} + \left[(1-2M_1 I_{xx})^2 + 2\omega_T^2 I_{xx}^2\right]^{\frac{1}{2}}\right\}, \tag{50}$$

$$\sigma_2 = \frac{1}{2I_{xx}}\left\{1+2M_1 I_{xx} - \left[(1-2M_1 I_{xx})^2 + 2\omega_T^2 I_{xx}^2\right]^{\frac{1}{2}}\right\}, \tag{51}$$

$$\sigma_3 = \frac{1}{2I_{zz}}\left\{1+2M_3 I_{zz} + \left[(1-2M_3 I_{zz})^2 + 2\omega_T^2 I_{zz}^2\right]^{\frac{1}{2}}\right\}, \tag{52}$$

$$\sigma_4 = \frac{1}{2I_{zz}}\left\{1+2M_3 I_{zz} - \left[(1-2M_3 I_{zz})^2 + 2\omega_T^2 I_{zz}^2\right]^{\frac{1}{2}}\right\}. \tag{53}$$

The six zero eigenvalues are associated with the six-dimensional complement space of the tangent space to the invariant manifold at the equilibrium, and the remaining six eigenvalues are associated with the six-dimensional tangent space to the invariant manifold $T\Sigma|_{z_e}$. Therefore, the conditions of nonlinear stability are that all the remaining six eigenvalues are positive. Notice that $1/I_{yy}$ is always positive, $\sigma_2 > 0$ implies $\sigma_1 > 0$, and $\sigma_4 > 0$ implies $\sigma_3 > 0$, we will have the conditions of nonlinear stability as follows:

$$M_2 > 0, \sigma_2 > 0, \sigma_4 > 0. \tag{54}$$

According to Eqs. (51) and (53), conditions of nonlinear stability Eq. (54) can be written as

$$M_2 > 0, 4M_1 > \omega_T^2 I_{xx}, 4M_3 > \omega_T^2 I_{zz}. \tag{55}$$

According to Eqs. (46)-(48), we can write conditions of nonlinear stability Eq. (55) further as

$$\left(\frac{3\mu}{R_S^3}-\frac{15\mu\tau_0}{2R_S^5}-\frac{57\mu\tau_2}{R_S^5}\right)(I_{xx}-I_{zz}) > 0, \tag{56a}$$

$$\left(\frac{6\mu\tau_2}{R_S^5}-\frac{3\mu\tau_0}{R_S^5}+\omega_T^2\right)(I_{yy}-I_{xx}) > 0, \tag{56b}$$

$$\left(\frac{3\mu}{R_S^3}-\frac{21\mu\tau_0}{2R_S^5}-\frac{51\mu\tau_2}{R_S^5}+\omega_T^2\right)(I_{yy}-I_{zz}) > 0. \tag{56c}$$

According to Eq. (4), we have

$$\omega_T^2 = \frac{\mu}{R_s^3} - \frac{\mu}{R_s^5}\left(\frac{3}{2}\tau_0 + 9\tau_2\right). \tag{57}$$

Then, Eqs. (56b) and (56c) can be written as follows:

$$\left(\frac{\mu}{R_S^3} - \frac{9\mu\tau_0}{2R_S^5} - \frac{3\mu\tau_2}{R_S^5}\right)(I_{yy} - I_{xx}) > 0, \tag{58a}$$

$$\left(\frac{\mu}{R_S^3} - \frac{3\mu\tau_0}{R_S^5} - \frac{15\mu\tau_2}{R_S^5}\right)(I_{yy} - I_{zz}) > 0. \tag{58b}$$

Keeping in mind that $\tau_0 = a_e^2 C_{20}$ and $\tau_2 = a_e^2 C_{22}$, we can write the stability conditions Eqs. (56a), (58a) and (58b) as follows:

$$\left[1 - \left(\frac{a_e}{R_S}\right)^2 \left(\frac{5}{2}C_{20} + 19C_{22}\right)\right](I_{xx} - I_{zz}) > 0, \tag{59a}$$

$$\left[1 - \left(\frac{a_e}{R_S}\right)^2 \left(\frac{9}{2}C_{20} + 3C_{22}\right)\right](I_{yy} - I_{xx}) > 0, \tag{59b}$$

$$\left[1 - \left(\frac{a_e}{R_S}\right)^2 (3C_{20} + 15C_{22})\right](I_{yy} - I_{zz}) > 0. \tag{59c}$$

The conditions of nonlinear stability Eqs. (59a)-(59c) can be rearranged further as follows:

$$A_{ast}\sigma_x > 0, \ B_{ast}\sigma_y > 0, \ C_{ast}\sigma_z > 0, \tag{60}$$

where $\sigma_x$, $\sigma_y$ and $\sigma_z$ are defined as

$$\sigma_x = \left(\frac{I_{yy} - I_{zz}}{I_{xx}}\right), \tag{61a}$$

$$\sigma_y = \left(\frac{I_{xx} - I_{zz}}{I_{yy}}\right), \tag{61b}$$

$$\sigma_z = \left(\frac{I_{yy} - I_{xx}}{I_{zz}}\right). \tag{61c}$$

and the parameters $A_{ast}$, $B_{ast}$ and $C_{ast}$ are defined by

$$A_{ast} = 1 - 3\left(\frac{a_e}{R_s}\right)^2 (C_{20} + 5C_{22}), \tag{62a}$$

$$B_{ast} = 1 - \left(\frac{a_e}{R_s}\right)^2 \left(\frac{5}{2}C_{20} + 19C_{22}\right), \tag{62b}$$

$$C_{ast} = 1 - \left(\frac{a_e}{R_s}\right)^2 \left(\frac{9}{2}C_{20} + 3C_{22}\right). \tag{62c}$$

The parameters $A_{ast}$, $B_{ast}$ and $C_{ast}$ are defined same as in Reference [20]. Obviously, the ranges of $\sigma_x$, $\sigma_y$ and $\sigma_z$ are all from -1 to 1. Taking into account that $\sigma_y > 0$ is equivalent to $\sigma_x > \sigma_z$, we can write the conditions of nonlinear stability Eq. (60) as follows:

$$A_{ast}\sigma_x > 0, \ B_{ast}(\sigma_x - \sigma_z) > 0, \ C_{ast}\sigma_z > 0. \tag{63}$$

When the problem is reduced to the attitude motion on a circular orbit in a central gravity field, we have $A_{ast} = B_{ast} = C_{ast} = 1$. Then, Eq. (63) is reduced to

$$\sigma_x > 0, \sigma_x > \sigma_z, \sigma_z > 0. \tag{64}$$

Here we have obtained the classical Lagrange region, which has already been obtained by Hughes[3], Beck and Hall[24] in the studies on the nonlinear attitude stability on a circular orbit in a central gravity field. The differences between our results Eq. (63) and the classical results Eq. (64) are due to the parameters $A_{ast}$, $B_{ast}$ and $C_{ast}$, i.e. the non-central gravity field of the asteroid.

**Some Discussions on the Parameters**

According to Eq. (63), the signs of functions $A_{ast}$, $B_{ast}$ and $C_{ast}$ have important qualitative effects on the conditions of nonlinear stability. According to Eqs. (62a)-(62c), the functions $A_{ast}$, $B_{ast}$ and $C_{ast}$ are determined by three basic parameters of the asteroid, including the ratio of the mean radius to the stationary orbital radius $a_e/R_s$, the harmonic coefficients $C_{20}$ and $C_{22}$.

Precisely speaking, the ratio $a_e/R_s$ depends on the harmonic coefficients $C_{20}$ and $C_{22}$, the average density and the rotational period of the asteroid. However, from an approximate point of view, the parameter $a_e/R_s$ can be roughly determined by the average density and the rotation period of the asteroid, with the effects of the harmonic coefficients $C_{20}$ and $C_{22}$ neglected. Wang and Xu[20] have made a rough estimate of the range of the ratio $a_e/R_s$, and have shown that the range from 0.2 to 0.8 should cover most asteroids in our Solar System. The ratio $a_e/R_s$ will be treated as the third parameter of the asteroid in the conditions of nonlinear stability described by Eq. (63), besides the harmonic coefficients $C_{20}$ and $C_{22}$. Therefore, the practical ranges of the three parameters in Eq. (63) are as follows:

$$\begin{cases} 0.2 < \dfrac{a_e}{R_s} < 0.8, \\ -0.5 < C_{20} < 0, \\ -0.25 < C_{22} < 0.25. \end{cases} \tag{65}$$

In a similar manner to Reference [20], we can divide the range of the parameter $a_e/R_s$ into three parts as follows:

(I). $$0.2 < \frac{a_e}{R_s} \leq \frac{2}{\sqrt{19}}, \tag{66}$$

in the case of which the functions $A_{ast}$, $B_{ast}$ and $C_{ast}$ are all positive in the domain of the harmonic coefficients given by $-0.5 < C_{20} < 0$, $-0.25 < C_{22} < 0.25$;

(II).
$$\frac{2}{\sqrt{19}} < \frac{a_e}{R_s} \leq \frac{2}{\sqrt{15}}, \tag{67}$$

in the case of which the functions $A_{ast}$ and $C_{ast}$ are both positive, but $B_{ast}$ can be positive or negative in the domain $-0.5 < C_{20} < 0$, $-0.25 < C_{22} < 0.25$;

(III).
$$\frac{2}{\sqrt{15}} < \frac{a_e}{R_s} < 0.8, \tag{68}$$

in the case of which the functions $C_{ast}$ is positive, but $A_{ast}$ and $B_{ast}$ can be positive or negative in the domain $-0.5 < C_{20} < 0$, $-0.25 < C_{22} < 0.25$.

## NONLINEAR STABILITY DOMAIN IN THE $\sigma_x$-$\sigma_z$ PLANE

The domain of harmonic coefficients $-0.5 < C_{20} < 0$, $-0.25 < C_{22} < 0.25$ can be divided into four parts according to the signs of the functions $A_{ast}$, $B_{ast}$ and $D_{ast}$, where $D_{ast}$ is defined as

$$D_{ast} = 1 - \left(\frac{a_e}{R_s}\right)^2 \left(\frac{3}{2}C_{20} + 9C_{22}\right). \tag{69}$$

According to Eq. (57), we have

$$\omega_T^2 = \frac{\mu}{R_s^3} D_{ast}. \tag{70}$$

Therefore, $D_{ast}$ should be positive, and the case $D_{ast} < 0$ does not exist in the real situation.

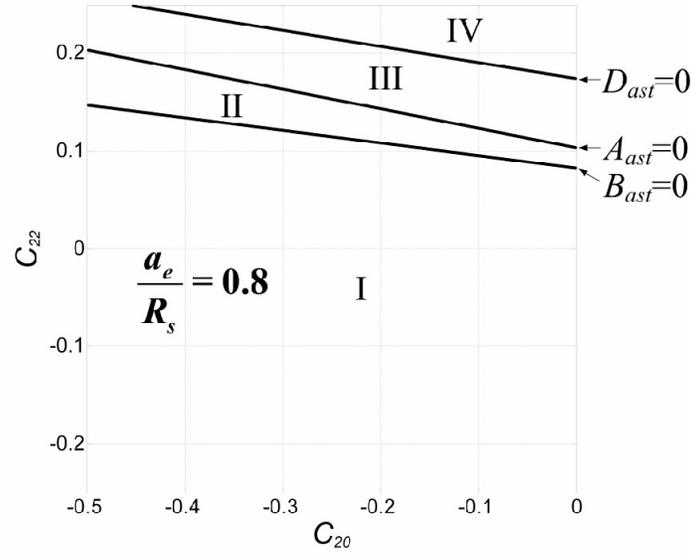

**Figure 2. The domain of the harmonic coefficients is divided into region I, region II, region III and region IV according to the sign of $A_{ast}$, $B_{ast}$ and $D_{ast}$ in the case of $a_e/R_s = 0.8$.**

Figure 2 has shown the regions I, II, III and IV in the case of $a_e/R_s = 0.8$. The straight lines $A_{ast} = 0$, $B_{ast} = 0$, $C_{ast} = 0$, $D_{ast} = 0$ on the $C_{20} - C_{22}$ plane are same as in Reference [20], therefore the regions I, II, III and IV are also same as in Reference [20]. According to Reference [20], the ratio $a_e/R_s$ has important effects on the conditions of the nonlinear stability. In the case of $0.2 < a_e/R_s \leq 2/\sqrt{19}$, only the region I can exist in the domain of the harmonic coefficients; the regions I and II can exist when $2/\sqrt{19} < a_e/R_s \leq 2/\sqrt{15}$; in the case of $2/\sqrt{15} < a_e/R_s \leq 2/3$, the regions I, II and III can exist; in the case of $2/3 < a_e/R_s \leq 0.8$, all the four regions can exist.

In the region I of the domain of the harmonic coefficients in Figure 2, the functions $A_{ast}$, $B_{ast}$ and $C_{ast}$ given by Eqs. (62a)-(62c) are all positive. Then, the conditions of nonlinear stability Eq. (63) can be written as Eq. (64), the classical nonlinear stability conditions on a circular orbit in a central gravity field. Therefore, in this case the nonlinear stability domain in the $\sigma_x$-$\sigma_z$ plane is the classical Lagrange region, same as the classical results on a circular orbit in a central gravity field, as shown by Figure 3.

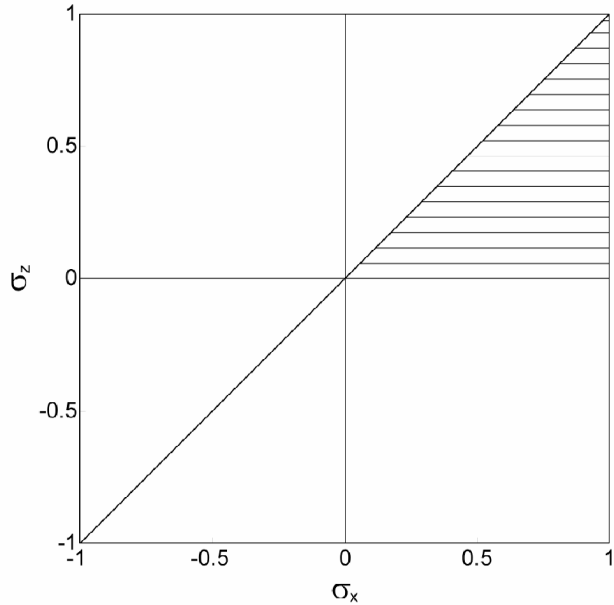

Figure 3. The nonlinear stability domain in the $\sigma_x$-$\sigma_z$ plane in the region I of Figure 2.

In the region II of the domain of the harmonic coefficients in Figure 2, the functions $A_{ast}$ and $C_{ast}$ are positive, and $B_{ast}$ is negative. Then, the conditions of nonlinear stability Eq. (63) can be written as

$$\sigma_x > 0, \; \sigma_x - \sigma_z < 0, \; \sigma_z > 0. \qquad (71)$$

Therefore, the nonlinear stability domain in the $\sigma_x$-$\sigma_z$ plane is an isosceles right triangle region above the straight line $\sigma_x - \sigma_z = 0$ in the I quadrant, as shown by Figure 4. Obviously, due to the non-spherical mass distribution of the asteroid, the stability domain is totally different from the classical results by Hughes[3], Beck and Hall[24] on a circular orbit in a central gravity field. This result is very important for the design of attitude control system of the asteroid missions.

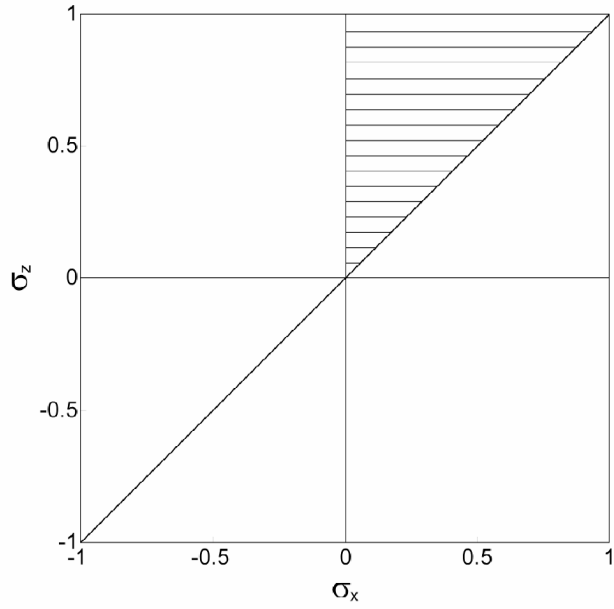

**Figure 4. The nonlinear stability domain in the $\sigma_x$-$\sigma_z$ plane in the region II of Figure 2.**

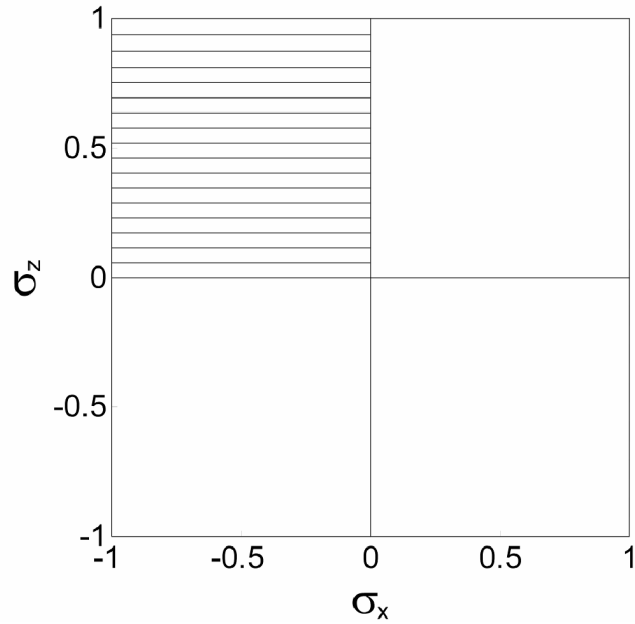

**Figure 5. The nonlinear stability domain in the $\sigma_x$-$\sigma_z$ plane in the region III of Figure 2.**

In the region III of the domain of the harmonic coefficients in Figure 2, the function $C_{ast}$ is positive, and $A_{ast}$ and $B_{ast}$ are negative. Then, the conditions of nonlinear stability Eq. (63) can be written as

$$\sigma_x < 0, \ \sigma_x - \sigma_z < 0, \ \sigma_z > 0. \tag{72}$$

Therefore, the nonlinear stability domain in the $\sigma_x$-$\sigma_z$ plane is the II quadrant, as shown by Figure 5. Due to the non-spherical mass distribution of the asteroid, this stability domain is also totally different from the classical results by Hughes[3], Beck and Hall[24] on a circular orbit in a central gravity field.

In the region IV of the domain of the harmonic coefficients in Figure 2, $D_{ast}$ is negative. As shown above, this case does not exist in the real physical situation.

Notice that the nonlinear attitude stability is more practical than the linear attitude stability studied in Reference [20]. Since the system is conservative and only the necessary conditions of stability can be obtained via the linearized equations of motion, the linear stability domain obtained there are only infinitesimally stable, but the stability can not be guaranteed for the finite motions; whereas the nonlinear attitude stability obtained in this paper can be guaranteed for the finite motions.

## CONCLUSION

The equilibrium attitude and the nonlinear stability of a rigid spacecraft on a stationary orbit around a uniformly-rotating asteroid have been studied in the framework of the geometric mechanics. In the studied problem, the harmonic coefficients $C_{20}$ and $C_{22}$ of the gravity field of the asteroid were considered. The tools of the geometric mechanics adopted in the paper provided a method for determining the equilibrium attitude from a global point of view and the energy-Casimir method for the conditions of the nonlinear stability.

Starting from the natural symplectic structure, we have derived the non-canonical Hamiltonian structure of the problem. The Poisson tensor, Casimir functions and equations of motion were obtained in a differential geometric method. 24 equilibrium attitudes of the spacecraft, which correspond to stationary points of the Hamiltonian constrained by Casimir functions, were determined from a global point of view.

The conditions of the nonlinear stability of the equilibrium attitude were obtained in a modified energy-Casimir method. Nonlinear stability of the equilibrium attitude was then investigated versus three basic parameters of the asteroid, including the ratio of the mean radius to the stationary orbital radius, the harmonic coefficients $C_{20}$ and $C_{22}$.

We have found that due to the significantly non-spherical shape and the rapid rotation of the asteroid, the nonlinear attitude stability domain in the $\sigma_x$-$\sigma_z$ plane is modified significantly in comparison with the classical nonlinear stability domain, i.e. the Lagrange region. In the different regions of the domain of the harmonic coefficients, the nonlinear stability properties of the equilibrium attitude can be totally different. Especially, when the spacecraft is located on the intermediate-moment principal axis of the asteroid, i.e. $C_{22} > 0$, the nonlinear stability domain in the $\sigma_x$-$\sigma_z$ plane can be an isosceles right triangle region above the straight line $\sigma_x - \sigma_z = 0$ in the I quadrant or the II quadrant, totally different from the classical Lagrange region on a circular orbit in a central gravity field.

Our results are very useful for the design of attitude control system in the future asteroid missions.

## ACKNOWLEDGMENTS

This work is supported by the Innovation Foundation of BUAA for PhD Graduates.